\documentclass[manuscript]{aastex}

\usepackage{graphicx}
\usepackage[colorlinks=True,citecolor=blue,breaklinks=True]{hyperref}
\usepackage{pbox}
\usepackage{lscape}
\usepackage{pdflscape}
\usepackage{tabularx}
\usepackage{rotating}
\usepackage{color, colortbl}
\definecolor{Gray}{gray}{0.9}
\definecolor{LightCyan}{rgb}{0.88,1,1}

\shorttitle{CALLISTO Spectrometer at IISER-Pune}
\shortauthors{Sasikumar Raja et al.}

\begin{document}

\title{CALLISTO Spectrometer at IISER-Pune}

\author{K. Sasikumar Raja\altaffilmark{1}}
\affil{Indian Institute of Science Education and Research, Pashan, Pune - 411 008, India}
\email{sasikumar@iiserpune.ac.in}

\author{Prasad Subramanian\altaffilmark{1}}
\affil{Indian Institute of Science Education and Research, Pashan, Pune - 411 008, India}

\author{S. Ananthakrishnan\altaffilmark{2}}
\affil{Department of Electronic Science, University of Pune, Pune - 411 007, India.}

\author{Christian Monstein\altaffilmark{3}}
\affil{Institute​ for​ Particle​ Physics​ and​ Astrophysics​, ETH​ Zurich,​ Switzerland.}

\altaffiltext{1}{Indian Institute of Science Education and Research, Pashan, Pune - 411 008, India} 
\altaffiltext{2}{Department of Electronic Science, University of Pune, Pune - 411 007, India.}
\altaffiltext{3}{Institute​ for​ Particle​ Physics​ and​ Astrophysics​, ETH​ Zurich,​ Switzerland.}

\begin{abstract}
A CALLISTO spectrometer to monitor solar radio transient emissions from $\approx 0.8-1.6~R_{\odot}$ (above photosphere)
is installed at IISER, Pune, India (longitude $73^{\circ} 55^{\arcmin}$ E and latitude $18^{\circ}31^{\arcmin}$ N). In this paper, 
we illustrate the instrumental details (log-periodic dipole antenna and the receiver system) along 
with the recorded solar radio bursts and 
radio frequency interferences produced by the thunderstorms in the frequency range 45-870 MHz.
We also developed the image processing pipelines using `sunpy' and in-house developed python library called `pycallisto'.
\end{abstract}


\section{Introduction}\label{intro}
A Compound Astronomical Low-Cost Low-Frequency Instrument for Spectroscopy and Transportable Observatory 
(CALLISTO) is radio spectrometer which is primarily designed to observe the transient radio emissions / 
radio bursts from the solar corona \citep{Ben2005, Ben2009}. The dominant magnetic field controls the structure evolution, 
kinematics and dynamics in the solar corona. The Sun is said to be `undisturbed' or quiet if there are no 
contribution from the transient events (sunspots, plages, active regions, filaments, prominences and other
transient eruptions like solar flares). The slowly varying component is observed above and near the vicinity 
of the active regions and chromospheric plages \citep{Kun1965}. These two emissions are thermal in nature. Presence of the 
previously mentioned solar events along with the large scale structures like Coronal Mass Ejections (CME) trigger 
the non-thermal radio emissions or radio bursts. In general, these bursts are classified into different types viz. Type I to Type V, based 
on drifting speed and morphology in the dynamic spectrum. Type J and 
Type U bursts are few other complex bursts which are often observed \citep{Kun1965, Mcl1985, Wil1967}. For the sake of completeness, 
we summarize properties of different radio bursts in a 
Table \ref{tab:rb}. In this paper, we describe solar radio Type II burst observed at IISER Pune on 04 November 2015. 

In order to monitor these bursts continuously (24 hours a day), CALLISTO spectrometers are distributed around the 
globe and established an e-CALLISTO network\footnote{http://www.e-callisto.org}. The real time data observed at different observatories are
uploaded to the server located at ETH Zurich, Switzerland automatically. Although, the instrument is designed to monitor the solar radio bursts, 
it is efficient to record the radio frequency interferences (RFI). Primarily, radio interferences or noise can be classified 
into two categories: (1) Man made or terrestrial interferences: modern life style demand the use of FM, AM, television, 
walkie-talkies, GSM and CDMA mobiles and are the sources of terrestrial interference. The industrial noise from automobiles,
aircraft ignition, electrical motors, switching equipment, voltage leakage from the electrical lines, fluorescent lights etc are 
few more examples. (2) Natural interferences: solar noise due to the activity in the Sun, cosmic noise from the celestial 
radio sources and transients, thunderstorms and lightening from Earth's atmosphere (also observed from other planets) and other
natural electrical discharges \citep{Bia2007, Ken2013}. In this paper, we also report the RFI produced during the thunderstorms. 

Since 2015, CALLISTO spectrometer is in operational at Indian Institute of Science Education and Research, Pune, India. 
The longitude and latitude of the place is $73^{\circ} 55^{\arcmin}$ E and $18^{\circ}31^{\arcmin}$ N respectively and situated at an altitude of 
558 meters above the sea level. Although, the instrument is located in the middle of the city, the radio spectrum is clean above 
115 MHz with only standard terrestrial interferences. 

\begin{sidewaystable}
\linespread{1.0}\selectfont \centering
\begin{tabular}[b]{llllllllll} 

\hline \hline
{\bf S.} & {\bf Type of} & {\bf Event } & {\bf Associated} & {\bf EM }&{\bf Frequency}&{\bf Duration}&{\bf dcp } \\ 
{\bf No.} &{\bf burst} & {\bf characteristics} &{\bf phenomenon}&& {\bf bandwidth}& &{\bf (\%)} \\ 

\hline \hline
\\
\rowcolor{LightCyan}
1&Type I & \pbox{10cm}{Narrow band, short \\ duration spikes superposed \\ over a continuum emission} & 
\pbox{5cm}{Active regions,\\ flares, eruptive \\ prominences}& PE&  50-500 MHz &
\pbox{3cm}{Single burst$\approx 1 \rm s$ \\ Noise storm: \\ hours-days  } & \pbox{3cm}{$\approx 0.5-1$ } \\ \\

\rowcolor{Gray}
2&Type II & \pbox{10cm}{Slow drifting $\approx 1$ MHz s$^{-1}$; \\second harmonics occurs} & 
\pbox{3cm}{Flares, MHD shocks,\\ proton emissions}& PE&  20-150 MHz &
\pbox{3cm}{3-30 mins} & \pbox{3cm}{$\approx 0.5$ }\\ \\

\rowcolor{LightCyan}
3&Type III & \pbox{10cm}{Fast drifting \\ $\approx 100$ MHz s$^{-1}$; \\ occurs isolated, \\ in groups or storms \\ second harmonics are seen} & 
\pbox{3cm}{Active regions,\\ flares}& PE&  10 kHz-1 GHz &
\pbox{4cm}{isolated$\approx 1-3 $ s \\ groups $\approx 1-10 $ mins \\ storms $\approx $ mins-hours} &
\pbox{3cm}{F$\approx 0.5$ \\ H$\lesssim 0.3$ }\\ \\

\rowcolor{Gray}
4&Type IVs & \pbox{10cm}{Smoothly varying \\ broad band continuum} & 
\pbox{3cm}{Flares, proton \\emissions}& GS&  20 MHz-2 GHz &
\pbox{3cm}{Hours-days  } & \pbox{3cm}{$\approx 0.5$ }\\ \\

\rowcolor{LightCyan}
5&Type IVm & \pbox{10cm}{Smoothly varying \\ broad band continuum; \\ slow drifting} & 
\pbox{4cm}{Eruptive \\ prominences, \\ MHD shocks}& GS&  20-400 MHz &
\pbox{3cm}{30-120 mins} & \pbox{3cm}{Increases \\ from low to \\ $\approx 1$ }\\ \\

\rowcolor{Gray}
6&Type V & \pbox{10cm}{Smooth, short lived \\continuum emission;\\ always follow by Type III\\ groups/storms} & 
\pbox{3cm}{Active regions,\\ flares}& PE&  10-200 MHz &
\pbox{3cm}{1-3 mins  } & \pbox{3cm}{very low \\ ($<0.1$)}\\ \\

\hline \hline
\end{tabular}

\caption{Low frequency radio bursts and their physical properties..
EM: Emission mechanism, PE: Plasma emission, GS: Gyrosynchrotron emission, dcp: degree of circular polarization.
For more details refer \citet{Kun1965,Mcl1985,Sas2015} and \url{http://www.sws.bom.gov.au/World_Data_Centre/1/9/5}.
}
\label{tab:rb} 
\end{sidewaystable}

\section{Instrumentation}
In this section, we describe different sections of our CALLISTO spectrometer.
\subsection{Log-periodic dipole antenna}
In solar corona, most of the radiation at radio wavelengths originates due to the coherent plasma emission mechanism.
In such case, plasma frequency ($f_p$) is proportional to the electron density ($N_e$) of that height \citep{Mcl1985, Ryb1986, Gar2004}. 
Since the electron density in the corona decreases with the increasing radial height, different frequency of observations correspond 
to specific layers in solar corona.
Hence, low frequency observations probe outer layers in corona. Therefore, we need a broadband antenna to understand radio bursts
and their propagation. 

In general, an antenna whose bandwidth extends 40:1 or more, are referred to frequency independent antenna. Log-periodic dipole 
(LPD) is one of such frequency independent antenna and its designing aspects were 
discussed by Duhamel $\&$ Isbell in 1957 \citep{Bal2005, Car1961, Ham1957, Sas2013}. The input impedance 
of the LPD is a periodic function of logarithm of the frequency and hence the name `log-periodic dipole'. Other antenna characteristics like: radiation patterns, 
directivity, beamwidth and side lobe levels etc also undergoes similar variations \citep{Bal2005}. Mathematically it can be expressed as, 
\begin{equation}
log(f_{n+1}) = log(f_n) + log(1/k)
\end{equation}

where, `f' is the frequency, `k' is the scaling factor, and `n' indicates dipole number.

\begin{figure*}[!ht]
\centerline{\includegraphics[width=12cm]{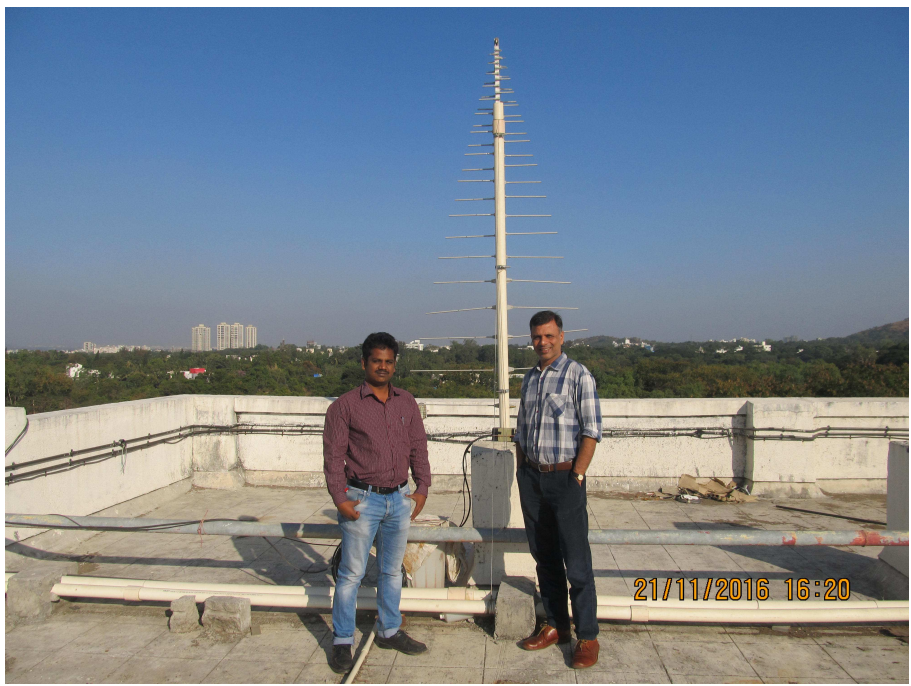}}
\caption{Log-periodic dipole antenna mounted at rooftop of the main building at IISER, 
Pune. The shorter arms of the LPD is pointing the zenith. The antenna arms are aligned 
along the North-South direction. With this setup, the Sun will be in the field of view 
(FOV) during 2-8 UT. As the LPD tilts by some angle, antenna beam also rotates
accordingly and therefore change in observing time. The broader E-plane of LPD enable us to monitor the 
Sun for all declinations of the year.
}
\label{fig:callisto_image}
\end{figure*}

The primary receiving element in the CALLISTO system is a LPD which operates over 100-900 MHz. Over the operating bandwidth the voltage standing wave ratio (VSWR) $\lesssim 2$. 
The LPD was designed and built in the Electronic Science Department of SP Pune University. It consists of a stack of dipoles (of different sizes)
mounted along two transmission lines (booms) in criss-cross fashion at different heights \citep{Ash2012}. The criss-cross fashion of the adjacent antenna arms 
introduce the phase shift of $180^{\circ}$ among them and cause the radiation to beam towards the shorter arms. Therefore, we should always mount the LPD such that shorter 
dipoles point toward the source direction (i.e. zenith in our case) as shown in Figure \ref{fig:callisto_image}. The signal from the LPD is tapped from top side using a co-axial cable (RG214) 
and run through the hallow square boom of the antenna in which the ground is connected. LPD is a linearly polarized antenna and have the half
power beamwidth (HPBW) in E-plane (along the direction of arms) $\approx 75^{\circ}$ and H-plane (perpendicular to the E-plane) 
$\approx 110^{\circ}$ \citep{Kis2014, Sas2015}. In the present case, 
we mounted LPD such that the dipoles are aligned parallel to the North-South direction. The broader HPBW of H-plane in East-West direction and E-plane in 
North-South direction allow us to observe the Sun within 2-8 UT and for all declinations of a year.

\subsection{CALLISTO Spectrometer}

\begin{figure*}[!ht]
\centerline{\includegraphics[width=12cm]{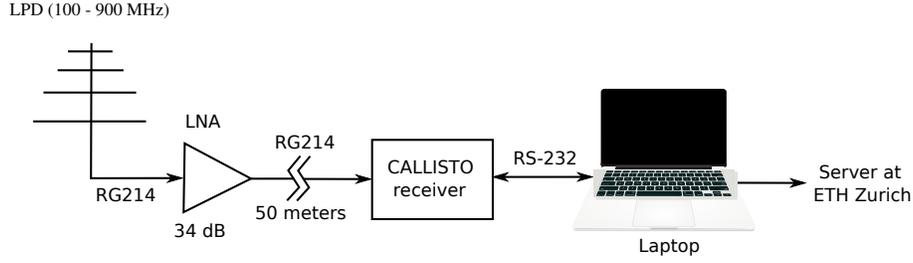}}
\caption{Schematic diagram of the CALLISTO setup. The broadband LPD output is pre-amplified with LNA 
of gain 34 dB at the rooftop. Using the RG214 coaxial cable the amplified signal is feed to the CALLISTO receiver. 
The spectrum is read to computer using RS-232 serial communication standard. The recorded data is fetched to the ftp-server at ETH Zurich.
}
\label{fig:drawing}
\end{figure*}

As shown in Figure \ref{fig:drawing}, the signal from the LPD is pre-amplified using a low noise amplifier (LNA) whose gain is 34 dB. 
The DC power supply of 12 V is supplied to the LNA from the receiver room. The LNA output is feed to the CALLISTO receiver using a RG214 
coaxial cable of length 50 meters. A linear DC power supply of 12 V (with current 225 mA) is supplied to the CALLISTO receiver. The operating 
bandwidth of the receiver is ≈ 45-870 MHz. This frequency range probes $\approx 0.8-1.6~R_{\odot}$ above photosphere ($1~R_{\odot} = 6.957 \times 10^8$ meters),
in the solar corona. The time and frequency resolutions 
are 250 ms (when channel number is 200) and 62.5 KHz respectively. The spectral data from the receiver is read to computer using RS-232 (serial 
communication standard) with a baud-rate of 115200 bits per second (bps). The data acquisition and other softwares are readily available on a 
CALLISTO web-page. Most of them are compatible with windows operating system. Every 15 minutes one file of size 760.3 kB will be saved into 
hard disk. A PERL script automatically uploads a newly created file to the ftp-server located at ETH Zurich, Switzerland and maintains a backup in the 
local machine. The data collected from different locations around the world forms a network called e-CALLISTO. The data is made accessible
to the international community\footnote{http://soleil.i4ds.ch/solarradio/callistoQuicklooks/}.

\subsection{Image Processing}

We processed the data stored in flexible image transport system (fits or fts) format recorded by the spectrometer using 
sunpy\footnote{http://docs.sunpy.org/en/stable/guide/tour.html?highlight=callisto}. Also, we developed a python library called `pycallisto',
which is available for the public at git-hub\footnote{https://github.com/ravipawase/pyCallisto}. The `pycallisto' library is developed using the standard python libraries: `matplotlib' and `pyfits'. 
The developed library can be used to process the data obtained at different observatories in addition to the non-CALLISTO spectrometers.

\section{Observations}

The preliminary observations carried out at IISER Pune were briefly discussed in this section. 

\subsection{Solar radio Type II bursts}

\begin{figure*}[!ht]
\centering
\begin{tabular}{cccc}
\includegraphics[scale=0.4]{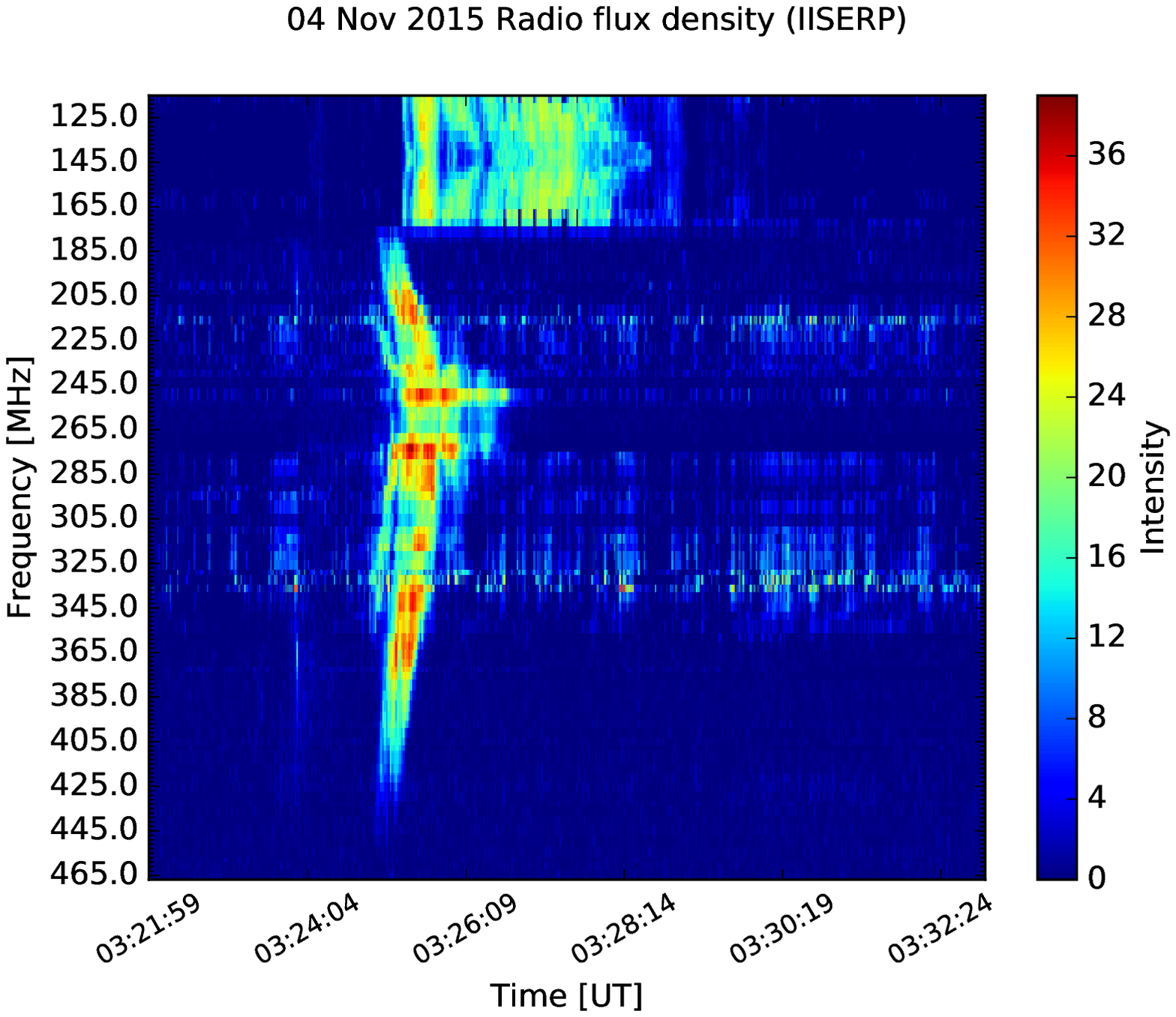} &
\includegraphics[scale=0.4]{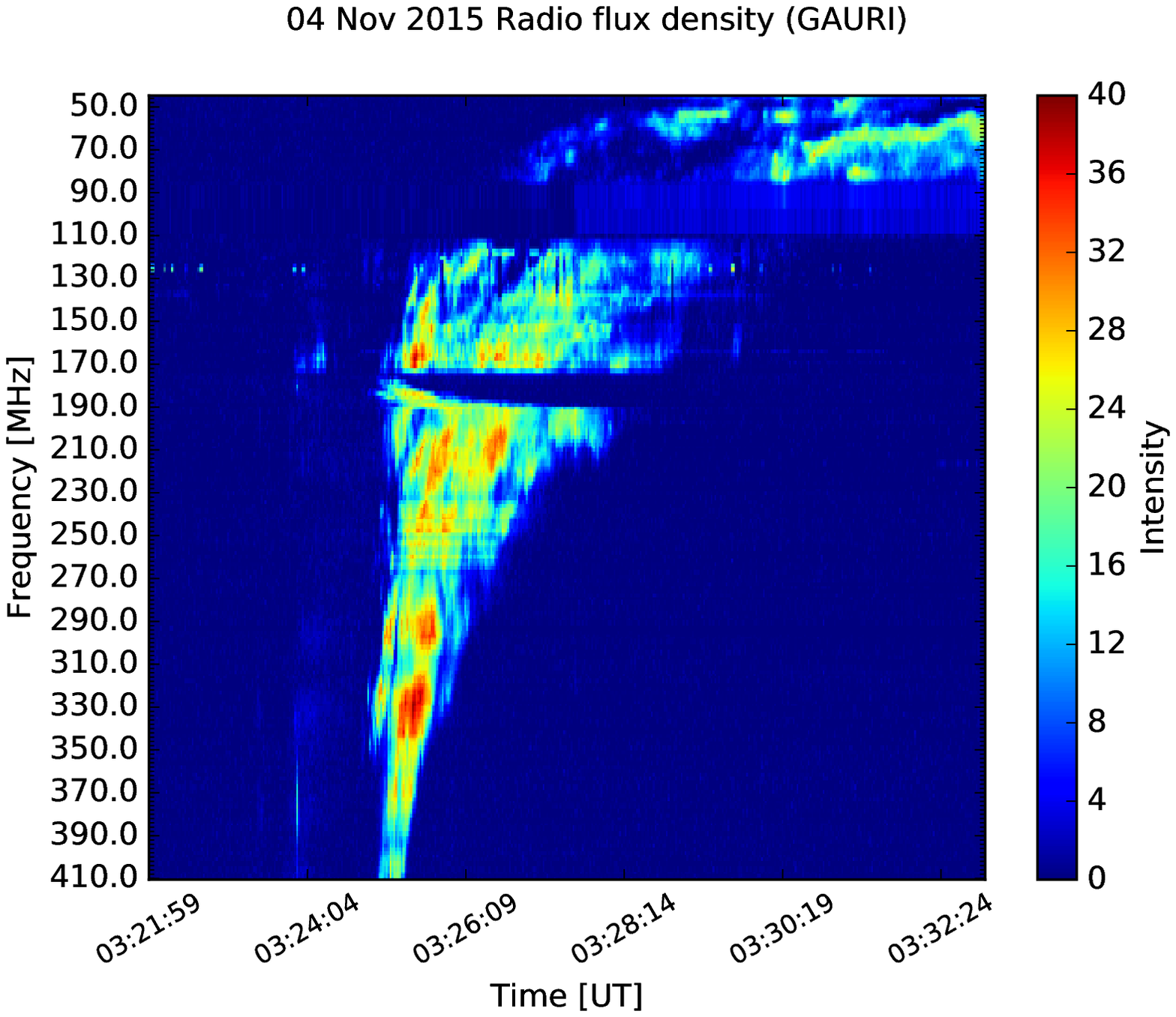} \\
\end{tabular}
\caption{The Figure shows the dynamic spectrum with solar Type radio II burst observed using 
CALLISTO at IISER Pune (on left-side) and at Gauribidanur observatory (right-side). 
This event was observed on 04 November 2015 at 03:23 UT. The burst was 
triggered by a M-class flare erupted at 03:20 UT. At IISER Pune and Gauribidanur Observatory the burst was observed 
in the frequencies range 115-465 MHz and 50-410 MHz respectively.}
\label{fig:type2}
\end{figure*}

Type II bursts are slow drifting bursts $\approx 1~ \rm MHz~ s^{-1}$. Type II bursts are the radio signatures of the shocks produced 
by a CME or flare. Since these bursts are electromagnetic in nature, they reach Earth in 8.3 minutes and provide 
the warning of interplanetary shocks \citep{Mcl1985,Gop2005}. Figure \ref{fig:type2} shows the first observation of Type II
radio burst observed using our CALLISTO spectrometer at IISER Pune (on left-side) and the same event observed at Gauribidanur observatory (on right-side) 
\citep{Ram2011} on 04 November 2015. The Type II radio burst was triggered by a shock caused by a reported M-class flare at 03:20 UT. 
The peak GOES X-ray flux of the flare was $\approx 1.9 \times 10^{-5}~ \rm W~m^{-2}$ at 03:25 UT\footnote{https://www.solarmonitor.org}. The flare was associated with an active region NOAA 12445 
located at heliographic coordinates N14W64. The Type II burst was first seen at 03:23 UT which lasted in 15 minutes in the 
frequency range 40-470 MHz. In the present case, the measured drifting speed of Type II burst was $\approx 1.18 ~\rm MHz~ s^{-1}$. 

\subsection{Lightning Signatures}
\begin{figure}[!ht]
\centerline{\includegraphics[width=10cm]{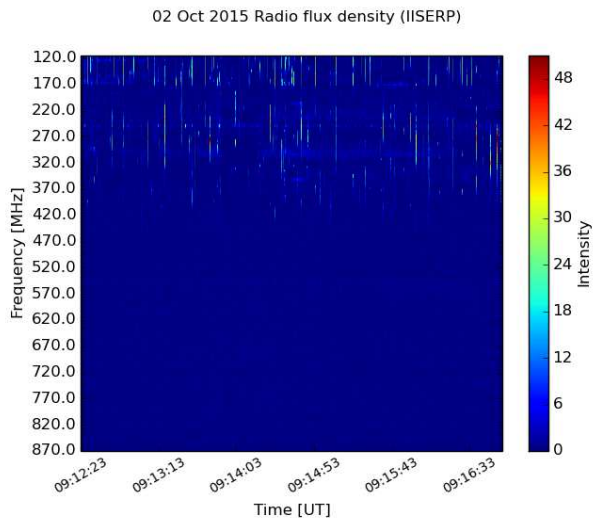}}
\caption{The dynamic spectrum shows the radio frequency interferences observed at IISER Pune 
during a thunderstorm in total power mode on 02 October 2015. Such observations are very common 
during monsoon with a sharp cutoff at $\approx 500$ MHz.
}
\label{fig:lightning}
\end{figure}

\begin{figure}[!ht]
\centerline{\includegraphics[width=10cm]{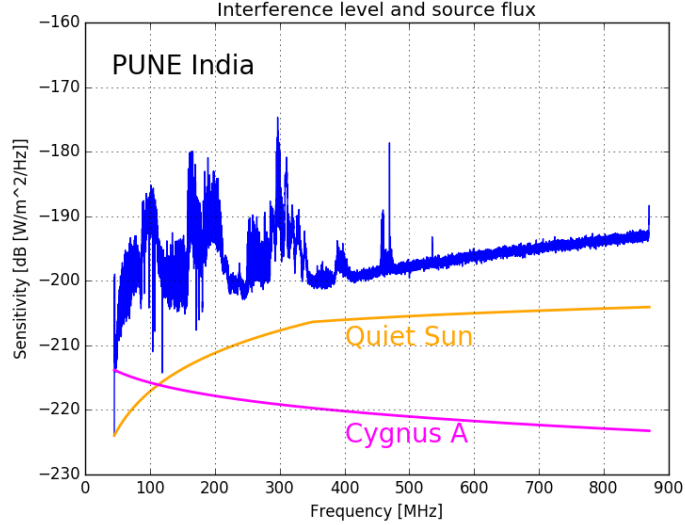}}
\caption{The figure shows the radio spectrum at IISER Pune. Note that the spectrum is recorded using CALLISTO 
(using spectral overview option) and hence harmonics of the FM are seen. 
However, the spectrum at higher frequencies is clean with a few selective interferences. The computed 
flux density of the quiet Sun and Cygnus-A is plotted for the reference.
}
\label{fig:rfi}
\end{figure}

As previously mentioned, there exists many natural and artificial sources which produce RFI. Thunderstorms (lightning)
produce RFI ranging from few kHz to MHz \citep{Smi1999, Sii2009}.
The Figure \ref{fig:lightning} shows the observed RFI due to the thunderstorms in the range 120-500 MHz on 02 October 
2015. During the monsoon, we have got many of such observations in 
the year 2015. Although, we verified the reasonable performance of the instrument in the frequency 
range 45-870 MHz by conducting few experiments, we noticed a sharp
cutoff of RFI above $\approx 500$ MHz. We have to investigate on it further. 
Figure \ref{fig:rfi} is the spectral overview recorded using CALLISTO at IISER Pune. Because of the closest 
FM stations many harmonics are seen in the spectrum.  Otherwise, the 
radio spectrum is clean at higher frequencies with a few selective interferences. 


\section{Summary and future plan}
Although, CALLISTO spectrometer is designed to monitor the corona continuously 24 hours a day by having a worldwide network called e-CALLISTO, 
it can be further used 
to monitor the terrestrial and natural interferences like thunderstorms. Note that observing thunderstorms may damage the whole setup including
computer by a lightning strike. It is advisable to use the lighting arresters for monitoring such interferences.
However, e-CALLISTO is a powerful network with more than 75 stations/observatories around the globe to monitor the solar radio bursts.
The e-CALLISTO network enable us to cross check the radio emissions across different observatories and to carry out scientific investigations. 
In future, at IISER Pune, we are planning to setup a spectropolarimeter. To achieve this, we are in a process of fabricating two more LPDs. 
By mounting these antennas in orthogonal fashion and using 4 port quadrature hybrid we can observe the left and right circularly polarized emissions \citep{Kis2015}. 
The quadrature hybrid outputs have to be connected to two separate CALLISTO receivers. Note that spectropolarimeter with increased sensitivity will play a 
crucial role in understanding the polarization properties of different radio bursts and magneto-ionic modes etc \citep{Rat1959, Asc2005}.

\section{Acknowledgements}
We thank the Electronic Science department of SP Pune University for long term 
loan of the LPD and the LNA. KSR acknowledges the financial support from the Science $\&$ 
Engineering Research Board (SERB), Department of Science $\&$ Technology, India (file: PDF/2015/000393). 
KSR would like to thank Nishtha Sachdeva, Tomin James, Spandan Choudhury, Amit Bhunia, 
Nilesh Dumbre and other staff members of IISER-Pune for their help in installation and maintenance of 
the instrument. KSR acknowledges Ravindra Pawase for working on image processing (python) library called `pycallisto'. 
\bibliographystyle{apj}
\bibliography{ms}

\end{document}